\documentclass[aps,prl,twocolumn,showpacs,10pt,superscriptaddress,preprintnumbers,nofootinbib,showkeys]{revtex4-1}
\usepackage{epsfig,amssymb,amsmath,psfrag,epstopdf,color}
\pdfoutput=1

\usepackage{graphicx}
\usepackage[margin=7pt,justification=centerlast]{caption}
\usepackage{subcaption}
\usepackage{slashed} 
\usepackage{paralist}
\usepackage{hyperref}
\usepackage[utf8]{inputenc}
\usepackage{booktabs}
\usepackage{bbold}

\DeclareMathOperator*{\SumInt}{%
\mathchoice%
  {\ooalign{$\displaystyle\sum$\cr\hidewidth$\displaystyle\int$\hidewidth\cr}}
  {\ooalign{\raisebox{.14\height}{\scalebox{.7}{$\textstyle\sum$}}\cr\hidewidth$\textstyle\int$\hidewidth\cr}}
  {\ooalign{\raisebox{.2\height}{\scalebox{.6}{$\scriptstyle\sum$}}\cr$\scriptstyle\int$\cr}}
  {\ooalign{\raisebox{.2\height}{\scalebox{.6}{$\scriptstyle\sum$}}\cr$\scriptstyle\int$\cr}}
}

\def\beq{\begin{equation}}
\def\eeq{\end{equation}}
\def\bsp#1\esp{\begin{split}#1\end{split}}
\newcommand{\be}{\begin{equation}}
\newcommand{\ee}{\end{equation}}
\newcommand{\bea}{\begin{eqnarray}}
\newcommand{\eea}{\end{eqnarray}}

\def\to{\rightarrow}

\def\ksl{\not{\hbox{\kern-2.3pt $k$}}}

\def\e{\epsilon}

\def\spa#1.#2{\left\langle#1\,#2\right\rangle}
\def\spb#1.#2{\left[#1\,#2\right]}
\def\lor#1.#2{\left(#1\,#2\right)}
\def\sand#1.#2.#3{%
\left\langle\smash{#1}{\vphantom1}^{-}\right|{#2}%
\left|\smash{#3}{\vphantom1}^{-}\right\rangle}

\newcommand{\nn}{\nonumber}

\newcommand{\mcdot}{\!\cdot\!}

\begin{document}

\title{Analytic Continuation and Reciprocity Relation for Collinear Splitting in QCD}
\author{Hao Chen}
\email{chenhao201224@zju.edu.cn}
\affiliation{Zhejiang Institute of Modern Physics, Department of
  Physics, Zhejiang University, Hangzhou, 310027, China\vspace{0.5ex}}
\author{Tong-Zhi Yang}
\email{yangtz@zju.edu.cn}
\affiliation{Zhejiang Institute of Modern Physics, Department of
  Physics, Zhejiang University, Hangzhou, 310027, China\vspace{0.5ex}}
\author{Hua Xing Zhu}
\email{zhuhx@zju.edu.cn}
\affiliation{Zhejiang Institute of Modern Physics, Department of
  Physics, Zhejiang University, Hangzhou, 310027, China\vspace{0.5ex}}
\author{Yu Jiao Zhu}
\email{zhuyujiao@zju.edu.cn}
\affiliation{Zhejiang Institute of Modern Physics, Department of
  Physics, Zhejiang University, Hangzhou, 310027, China\vspace{0.5ex}}

\begin{abstract}
It is well-known that direct analytic continuation of DGLAP evolution kernel~(splitting functions) from space-like to time-like kinematics breaks down at three loops. We identify the origin of this breakdown as splitting functions are not  analytic function of external momenta. However, splitting functions can be constructed from square of (generalized)~splitting amplitudes.
We establish the rule of analytic continuation for splitting amplitudes, and use them to determine the analytic continuation of certain holomorphic and anti-holomorphic part of splitting functions and transverse-momentum dependent distributions. In this way we derive the time-like splitting functions at three loops without ambiguity. We also propose a reciprocity relation for singlet splitting functions, and provide non-trivial evidence that it holds in QCD at least through three loops.
\end{abstract}


\maketitle

\section{Introduction}
\label{sec:introduction}

\begin{figure}[ht!]
  \begin{subfigure}[b]{0.4\linewidth}
    \includegraphics[width=1\textwidth]{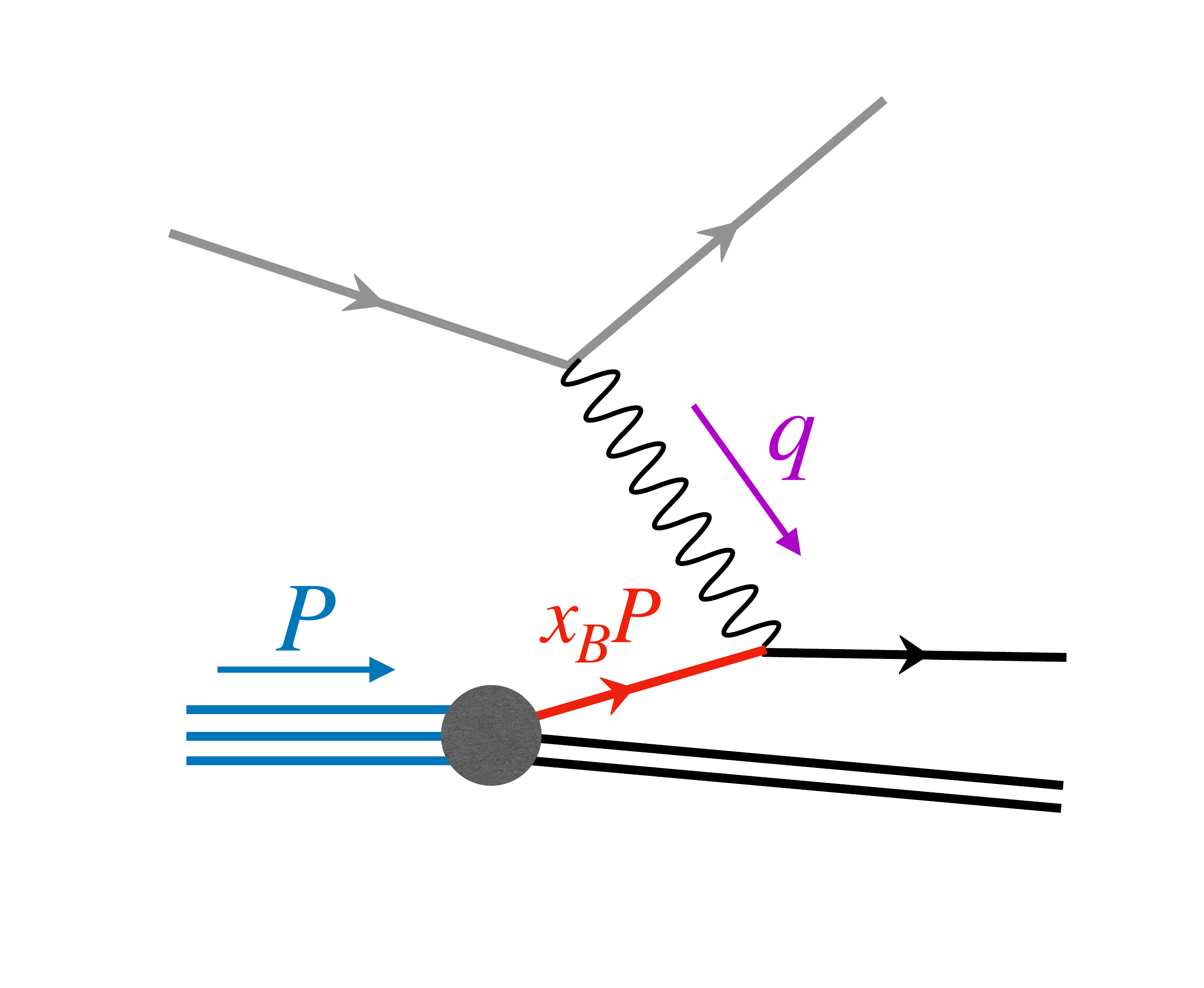}
    \caption{DIS}
\label{fig:1a}
  \end{subfigure}
  \begin{subfigure}[b]{0.4\linewidth}
    \includegraphics[width=1\textwidth]{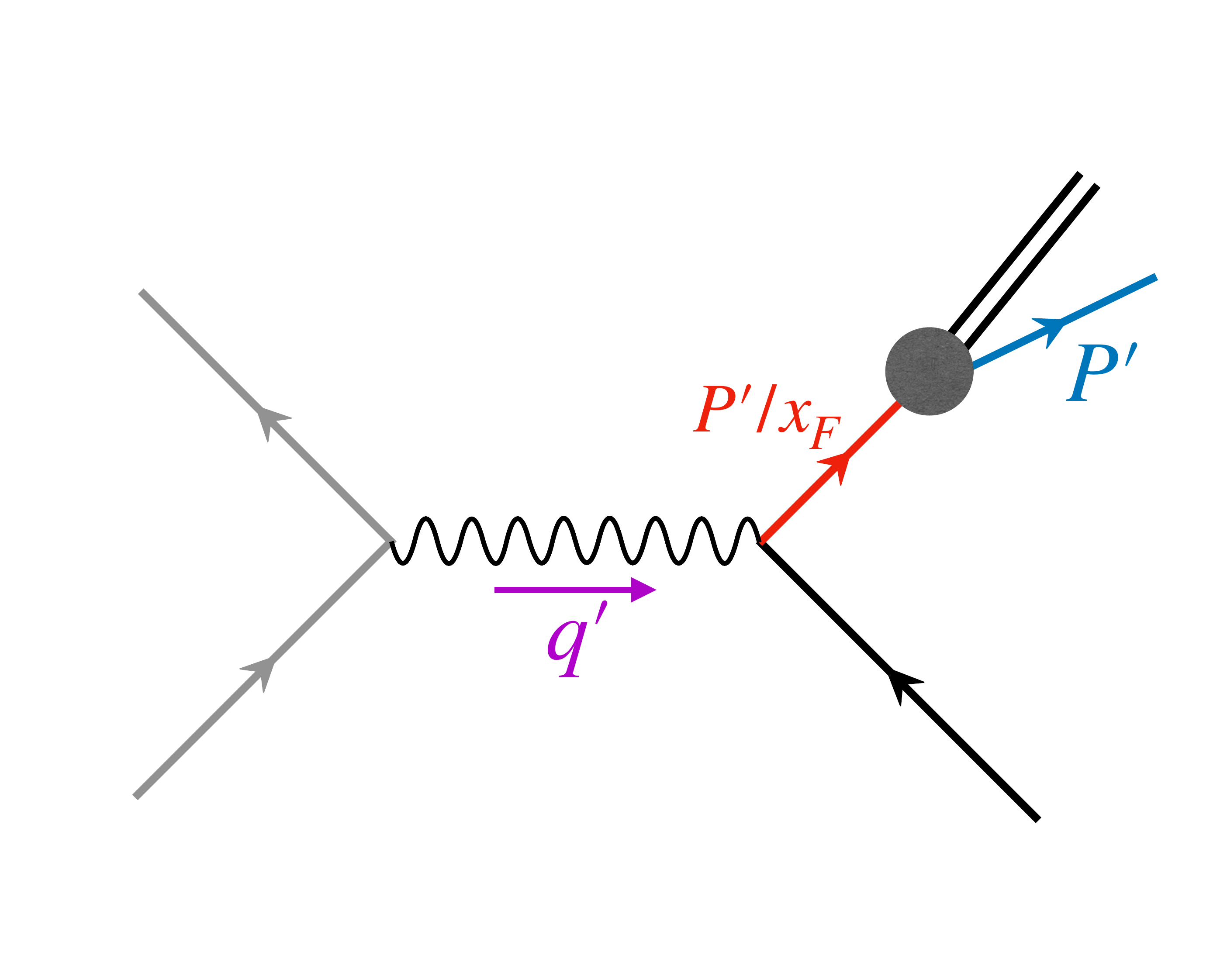}
    \caption{$e^+e^-$}
\label{fig:1b}
  \end{subfigure}
\caption{Typical processes used for the determination of PDFs~\ref{fig:1a} and FFs~\ref{fig:1b}.}
\label{fig:1}
\end{figure}

Parton Distributions Functions~(PDFs) and Fragmentation Functions~(FFs) provide essential input for accurate determination of various quantities of QCD and the Standard Model~\cite{Gao:2017yyd,deFlorian:2014xna,Anderle:2015lqa} within the framework of QCD factorization~\cite{Collins:1988ig}.   
While PDFs and FFs are intrinsically non-perturbative objects, their scale evolution obey Dokshitzer-Gribov-Lipatov-Altarelli-Parisi~(DGLAP) equations~\cite{Gribov:1972ri,Lipatov:1974qm,Altarelli:1977zs}. The corresponding evolution kernel are space-like~($q^2 < 0$, Fig.~\ref{fig:1a}) splitting functions for PDFs and time-like~($q'^2 > 0$, Fig.~\ref{fig:1b}) splitting functions for FFs, both of which can calculated in QCD perturbation theory. Determining the splitting functions to higher orders is one of the most important task of perturbative QCD.

Space-like splitting functions have been obtained to Next-to-Next-to-Leading Order~(NNLO) long time~\cite{Moch:2004pa,Vogt:2004mw}, and recently to N$^3$LO for the non-singlet ones~\cite{Moch:2017uml}. On the other hand, knowledge for time-like splitting functions are less precise. Direct calculation of time-like splitting functions have been done in \cite{Stratmann:1996hn} at NLO. At NNLO and beyond, results by direct calculation have not yet  been available~(see \cite{Gituliar:2015pra,Gituliar:2015iyq,Gituliar:2018bcr,Magerya:2019cvz} for recent progress). However, it has long been noted that the space-like Deep-inelastic Scattering and $e^+e^-$ annihilation are kinematically related~\cite{Drell:1969jm,Drell:1969wd}. The easiest way to see this is from the definition of Bjorken variable $x_B$ in DIS and Feynman variable $x_F$ in $e^+e^-$,
\begin{align}
x_B = \frac{-q^2}{2 P \cdot q} \,, \qquad x_F = \frac{2 P' \cdot q'}{q'^2}  \,,
\end{align}
where $P$ is the incoming hadron momenta in DIS, and $P'$ is the detected hadron momenta in $e^+e^-$, $q$ and $q'$ are the  space-like and time-like momentum transfer, respectively. After crossing, $P = - P'$, $q = q'$, one finds the analytic continuation relation $x_B=1/x_F$. However, beyond LO, the analytic continuation relation can not be applied directly to the splitting functions, but to appropriate bare quantities~\cite{Stratmann:1996hn,Blumlein:2000wh}. Analytic continuation of exclusive amplitudes has also been understood at NLO accuracy~\cite{Muller:2012yq}. Further subtleties arise at NNLO, where additional momentum sum rules, supersymmetry relation, and reciprocity consideration at large $x$~\cite{Dokshitzer:2005bf} are needed in order to obtain NNLO non-singlet and singlet time-like splitting functions~\cite{Mitov:2006ic,Moch:2007tx,Almasy:2011eq}. However, as have been explicitly pointed out in \cite{Almasy:2011eq}, the third order corrections to off-diagonal quark-gluon splitting, $P_{qg}^{\text{T},(2)}$, has only been determined up to an uncertainty proportional to QCD beta function. Fixing this remaining uncertainty is not only crucial for achieving complete NNLO analysis of parton-to-hadron fragmentation, but is also important for precision jet substructure study, see e.g.~\cite{Kang:2016ehg,Kang:2016mcy,Kang:2017frl,Larkoski:2017bvj,Gutierrez-Reyes:2018qez,Gutierrez-Reyes:2019vbx,Dixon:2019uzg,Chen:2020vvp}.

In this Letter we study the analytic continuation of splitting functions using Soft-Collinear Effective Theory~\cite{Bauer:2000ew,Bauer:2000yr,Bauer:2001yt,Bauer:2002nz}. We point out that splitting functions, both space-like or time-like, can be extracted from bare Transverse-Momentum-Dependent~(TMD) distributions. We identify the origin of the breakdown of direct analytic continuation for splitting functions and TMD distributions, as they are computed from square of splitting amplitudes, and therefore not analytic. Nevertheless, we identify certain holomorphic and anti-holomorphic contributions to TMD distributions, for which a correct rule of analytic continuation can be established. We use this to obtain time-like splitting functions at NNLO from the space-like ones.
Our results are in full agreement with those obtained in \cite{Mitov:2006ic,Moch:2007tx,Almasy:2011eq}, except a minor discrepancy in $P_{qg}^{\text{T},(2)}$. 
Finally, we propose an all-order generalization of Gribov-Lipatov reciprocity relation~\cite{Gribov:1972rt} for singlet splitting functions in QCD. Using the time-like splitting functions obtained in this work, we verify this relation to NNLO, where the discrepancy in $P_{qg}^{\text{T},(2)}$ mentioned above plays an important role.

\section{Splitting functions from TMD distributions}
\label{sec:splitt-funct-from}

TMD distributions are central ingredients in TMD factorization approach to hard scattering~\cite{Dokshitzer:1978yd,Parisi:1979se,Collins:1981uw,Collins:1984kg,Ji:2004xq,Ji:2004wu,Bozzi:2005wk,Cherednikov:2007tw,Collins:2011zzd,Becher:2010tm,Echevarria:2012pw,Chiu:2012ir}. In SCET, they can be conveniently defined as matrix element of collinear fields integrated over light-cone coordinate. Since for the purpose of analytic continuation, there is no intrinsic difference between quark and gluon TMD distributions, we shall focus on quark TMD distributions in the discussion below. The operator definition for quark TMD PDF is given by
\begin{multline}
  \mathcal{B}_{q/N}(x_B,b_\perp) = \SumInt_{X_n}
\int \frac{db^-}{2\pi} \, e^{-i x_B b^- P^+}  
\\ 
\cdot  \langle N(P) | \bar{\chi}_n(0,b^-,b_\perp)
| X_n \rangle \frac{\slashed{\bar{n}}}{2} 
\langle X_n| \chi_n(0) | N(P) \rangle \,,
\label{eq:TMDB}
\end{multline}
where $N(P)$ is a hadron state with momentum $P^\mu = (\bar{n} \mcdot P) n^\mu/2 = P^+ n^\mu/2$, with $n^\mu = (1, 0, 0, 1)$ and $\bar{n}^\mu = (1, 0, 0, -1)$. 
$\chi_n(x) = W^\dagger_n(x)  \xi_n(x)$ is gauge invariant collinear quark field~\cite{Bauer:2001ct}, and
\begin{gather}
W^\dagger_n(x)  = \mathcal{P} \exp{\left(ig_s \int_0^{\infty} ds \, \bar{n} \mcdot \boldsymbol{A}_n(x + s \bar{n} ) e^{-\varepsilon s}\right)} 
\end{gather}
is path-ordered $n$-collinear Wilson lines in fundamental representation. 
Although not necessary, we have inserted a complete set of $n$-collinear state $\mathbb{1} = \SumInt_{X_n} | X_n \rangle \langle X_n |$ into the definition of ${\cal B}_{q/N}$. Similarly, for an anti-quark $\bar{q}$ fragments into an anti-hadron $\overline{N}$, the TMD FF can be written as
\begin{multline}
\mathcal{F}_{\overline N/\bar q}(x_F,b_\perp)  = \SumInt_{X_n} x_F^{1-2 \epsilon} \int \frac{db^-}{2\pi}   e^{i b^-  P'^+ / x_F}  \\ 
\cdot  \langle 0 | 
 \bar \chi_{n}(0,b^-,b_\perp) | \overline N(P'),X_n \rangle \frac{\slashed{\bar{n}}}{2} \langle \overline N(P'),X_n | \chi_{n}(0) | 0 \rangle  \,,
\label{eq:TMDF}
\end{multline}
where $P'^\mu = (\bar n \mcdot P') n^\mu/2 =  P'^+ n^\mu/2$ is the momenta of the final state detected hadron. 
At high energy and small $|\vec{b}_\perp|$, TMD PDFs and FFs admit light-cone operator product expansion onto collinear PDFs and FFs, with perturbative calculable Wilson coefficients, which have been calculated to NNLO~\cite{Catani:2011kr,Catani:2012qa,Gehrmann:2012ze,Gehrmann:2014yya,Echevarria:2016scs,Luo:2019hmp,Luo:2019bmw,Gutierrez-Reyes:2019rug,Ebert:2020lxs}, and very recently to N$^3$LO~\cite{Luo:2019szz,Ebert:2020yqt}. The Wilson coefficients can be directly calculated by replacing the non-perturbative hadronic state $N\,(\overline{N})$ by perturbative partonic state $i\,(\bar \imath)$, namely ${\cal B}_{q/i}$ and ${\cal F}_{\bar \imath/\bar q}$. 
The operator definitions in Eq.~\eqref{eq:TMDB} and \eqref{eq:TMDF} make it clear that they can be computed from squared amplitudes integrated over collinear phase space~\cite{Ritzmann:2014mka}, 
\begin{align}
  {\cal B}_{q/i} = &\ \sum_{X_n} \int\! d {\rm PS}_{X_n} e^{- i K_\perp \!\cdot b_\perp} 
\delta(K^+ - (1 - x_B) P^+) 
\nn\\
&\ 
\cdot  {\rm \bold{Sp}}_{X_n q^* \leftarrow i}^{\rm S} \frac{\slashed{\bar{n}}}{2} 
{\rm \bold{Sp}}_{X_n q^* \leftarrow i}^{\rm S,*}  \,,
\nn
\\
{\cal F}_{\bar \imath/\bar q} = &\ \sum_{X_n}
x_F^{1 - 2 \e} \int \! d{\rm PS}_{X_n} 
e^{- i K_\perp \!\cdot b_\perp}  \delta\left(K^+ - \left(\frac{1}{x_F} - 1 \right) P'^+\right) 
\nn
\\
& \
\cdot
 {\rm \bold{Sp}}_{X_n \bar \imath \leftarrow \bar q^*}^{\rm T} 
\frac{\slashed{\bar{n}}}{2} 
 {\rm \bold{Sp}}_{X_n \bar \imath \leftarrow \bar q^*}^{\text{T},*} 
\,,
\label{eq:amp}
\end{align}
where $K^\mu$ is the total momentum of $|X_n \rangle$, $d\text{PS}_{X_n}$ is the collinear phase space measure. We have also defined the (generalized) space-like and time-like splitting amplitudes~\cite{Feige:2015rea,Schwartz:2017nmr}, 
\begin{align}
  {\rm \bold{Sp}}_{X_n q^* \leftarrow i}^{\rm S} \left(
k_a^+/P^+, \dots
 \right)  = &\ \langle X_n| \chi_n(0) | V_{P_l}^i(P_r) \rangle
\,, 
\nn\\
{\rm \bold{Sp}}_{X_n \bar \imath \leftarrow \bar q^*}^{\rm T} (k_a^+/P'^+, \dots)
= &\  \langle X_n, V_{P'_l}^{\bar \imath}(P'_r)  | \chi_{n}(0) | 0 \rangle \,, 
\label{eq:split}
\end{align}
where ${\rm \bold{Sp}}_{X_n q^* \leftarrow i}^{\rm S}$ denotes the amplitudes for parton $i$ splits into an off-shell quark $q^*$ and $X_n$, and similarly for ${\rm \bold{Sp}}_{X_n \bar \imath \leftarrow \bar q^*}^{\rm T}$.
$|V_{P_l}^i(P_r) \rangle$ denotes the partonic state $i$ with momentum $P$, decomposed into label momentum and residual momentum $P^\mu = P^\mu_l + P^\mu_r$, and similarly for $V_{P'_l}^{\bar \imath}(P'_r)$. Label momentum in SCET is Euclidean-like and do not require casual prescription, while residual momentum do and will be discussed in next section. When $X_n$ consists of a single parton, \eqref{eq:split} reduces to the usual $1\to 2$ splitting amplitudes, which are known to two-loop accuracy~\cite{Bern:2004cz,Badger:2004uk,Duhr:2014nda}. Results are also available for $1\to 3$ and $1\to 4$ splitting~\cite{Campbell:1997hg,Catani:1999ss,Badger:2015cxa,DelDuca:2019ggv}. In \eqref{eq:split} we have make explicit the possible functional dependence on $P^+$ and $P'^+$, where $k_a$ is any combination of momentum in $|X_n\rangle$. This is due to reparameterization III invariance in SCET~\cite{Manohar:2002fd}, namely SCET matrix element should be invariant under $n^\mu \to e^{\lambda} n^\mu$ and $\bar{n}^\mu \to e^{-\lambda} \bar{n}^\mu$. We have also made implicit in \eqref{eq:amp} average over initial-state spin and color, as well as sum over final-state spin and color.

After proper renormalization and zero-bin subtraction~\cite{Manohar:2006nz}, the TMD PDFs and FFs still contain collinear divergence due to the tagged hadron in initial state or final state. Schematically, at $n$-th order in perturbation theory, the single pole of the remaining collinear divergences have the following convolution form
\begin{align}
{\cal B}_{q/i}^{(n)} \sim \sum_j \frac{P_{qj}^{{\rm S}, (n)}}{n \e} \otimes \phi_{ji}^{\rm bare} 
\,, \quad
{\cal F}_{\bar \imath /\bar q}^{(n)} \sim \sum_j d_{\bar \imath j}^{\rm bare} \otimes \frac{P_{j \bar q}^{{\rm T},(n)}}{n \e} \,. \nn
\end{align}
where $\phi_{ij}^{\rm bare} = d_{ij}^{\rm bare} = \delta_{ij}$ are the bare partonic PDFs and FFs.
Therefore, one can extract the space-like and time-like splitting functions directly from the partonic TMD PDFs and FFs.

\section{Analytic Continuation of Splitting Amplitudes}
\label{sec:analyt-cont-s}

\begin{figure}[ht!]
  \centering
  \includegraphics[width=0.7\linewidth]{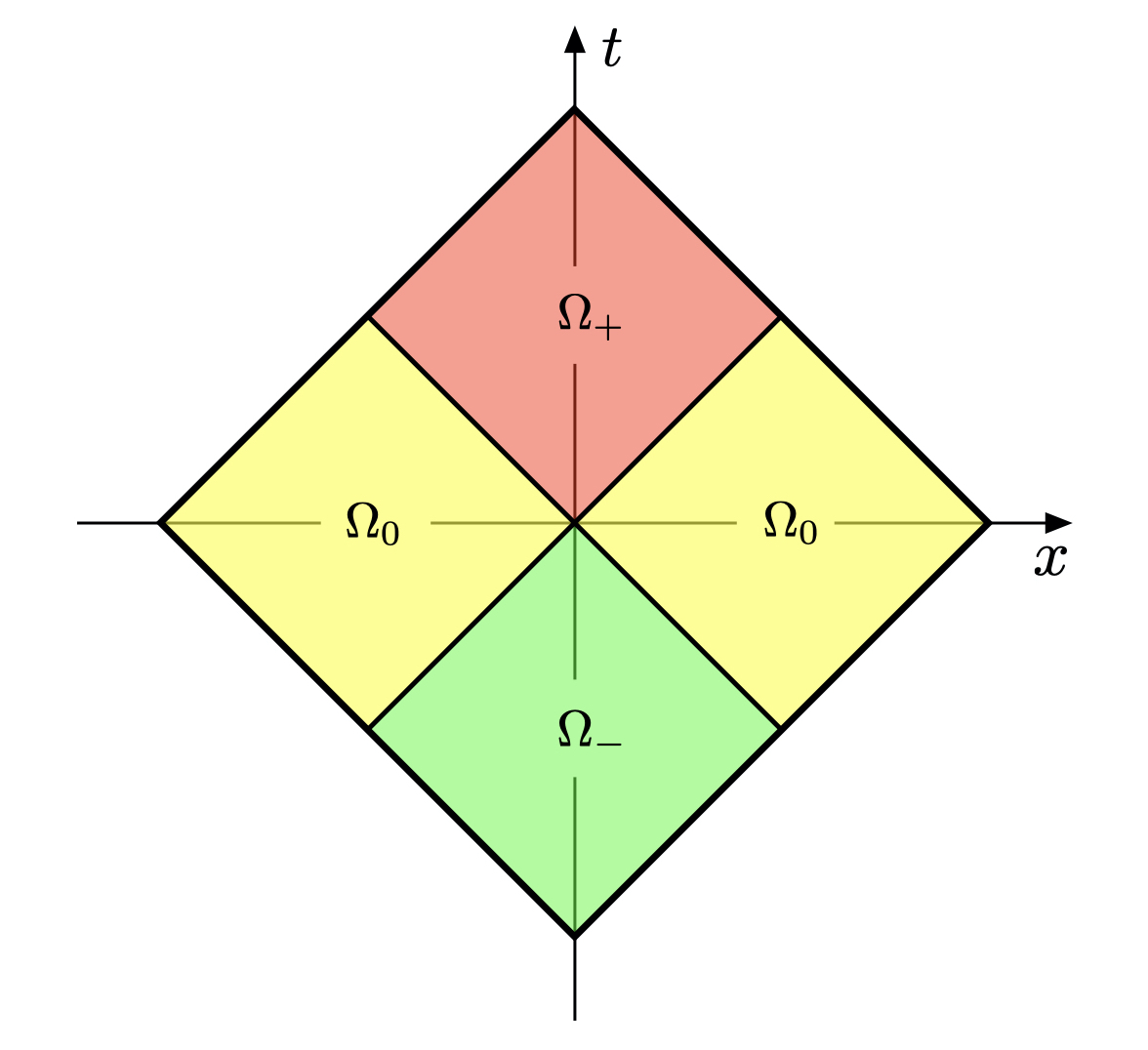}
  \caption{Penrose diagram of Minkowski space.}
  \label{fig:2}
\end{figure}

In order to understand the analytic continuation for TMD PDFs and FFs,
we start with LSZ reduction on the space-like splitting amplitudes:
\begin{align}
  {\rm \bold{Sp}}_{X_n q^* \leftarrow i}^{\rm S} \!
= \int\! d^dx \, e^{-i P_r \cdot x} 
\langle X_n |{\rm T} \{ \chi_n (0) J_{P_l}^i(x) \} | 0 \rangle \,,
\end{align}
where the current $J^{i}_{P_l}(x)\! =\!i(i\mathcal{P}_l+\partial_x)^2 V_{P_l}(x)$ creates a parton state $i$ from vacuum. Using that the SCET operator $\chi_n(x)$ is local in residual space, the time-ordering product be replaced by a (anti-)commutator if $i$ is a boson~(fermion),
\begin{align}
{\rm T}\{\chi_n(0)J^{i}_{P_l}(x)\}
=\theta(-x^0)\left[\chi_n(0),J^{i}_{P_l}(x)\right]_{\mp} \pm J^{i}_{P_l}(x)\chi_n(0) \,.
\label{eq:timeorder}
\end{align}
The second term in \eqref{eq:timeorder} doesn't contribute to the correlator since $\chi_n(0)$ effectively carries negative energy in the physical process and thus annihilate vacuum $| 0 \rangle$. Note that since $\chi_n$ is local, the (anti-)commutator in \eqref{eq:timeorder} vanishes in space-like region $\Omega_0$ of Fig.~\ref{fig:2}.
Thus, we can rewrite space-like splitting amplitudes as
\begin{align}
  {\rm \bold{Sp}}_{X_n q \leftarrow i}^{\rm S} \!=\! \int \limits_{x \in \Omega_-}\!\! d^dx \, e^{-i P_r \cdot x} 
\langle X_n | [ \chi(0), J_{P_l}^i(x) ]_{\mp} | 0 \rangle \,,
\end{align} 
where the $x$ integral is now restricted to inside the past light-cone, $\Omega_-$. 
Demanding analyticity for the splitting amplitudes imposes a unique casual prescription for residual momenta,  $P_r\to P_r+i q_I$ where $q_I$ is any positive-energy time-like vector.

Similarly, we can write time-like splitting amplitudes as
\begin{align}
{\rm \bold{Sp}}_{X_n \bar \imath \leftarrow \bar q}^{\rm T}
= \int\limits_{x \in \Omega_+} \!\!
d^d x \, e^{i P'_r \cdot x}  
\langle X_n | [ J_{P'_l}^{\bar \imath}(x) , \chi(0) ]_{\mp} | 0 \rangle \,,
\end{align}
and again the casual prescription must be $P^\prime_r \to P^\prime_r + i q_I$. With the casual prescription properly defined, we can now discuss the analytic continuation between space-like and time-like splitting amplitudes.

Since splitting amplitudes are analytic functions of external momentum, we can continue $P$ and $P'$ to common space-like infinity region, where space-like and time-like splitting amplitudes can be shown to equal. 
Therefore, by edge-of-the-wedge theorem~\cite{Bogolyubov:1956bzj}, space-like and time-like splitting amplitudes are actually analytic continuation of each other, although the casual prescription described above tell us their analytic region are disjoint.

\begin{figure}[ht!]
  \centering
  \includegraphics[width=1\linewidth]{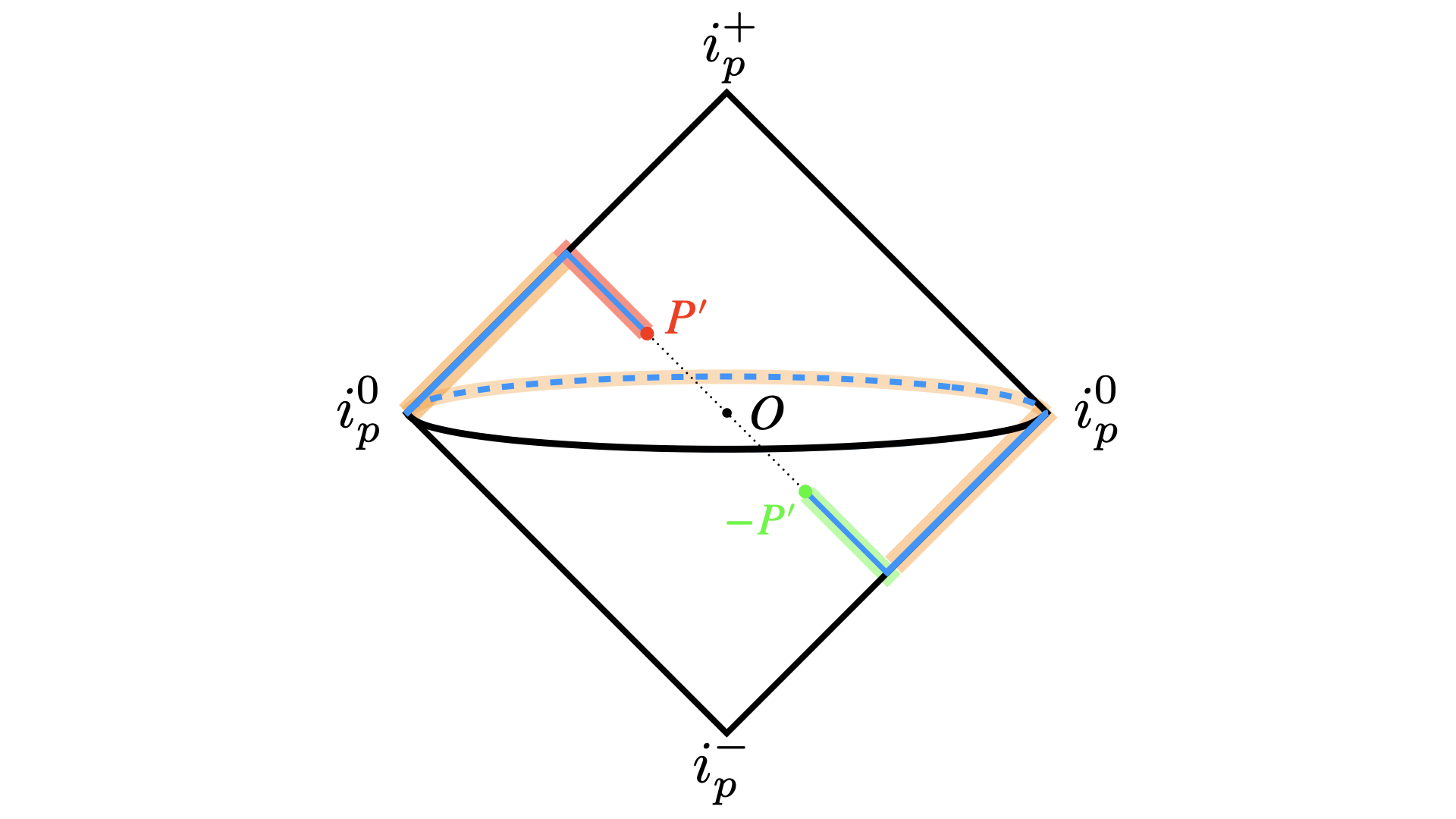}
  \caption{Penrose diagram of real momentum space. We have shown an extra spatial momentum dimension to visualize the path of analytic continuation, the blue lines.}
  \label{fig:3}
\end{figure}

For concreteness and later convenience, we choose a particular path displayed in Fig.~(\ref{fig:3}) as the blue lines (solid and dash), where we analytic continue the momentum of a time-like splitting amplitude from $P'$~(red) to $-P'$~(green). The orange segment of the path, sitting at space-like infinity relative to $O$, lies inside the region where ${\rm \bold{Sp}}^{\rm S}(-P^\prime)\!=\!{\rm \bold{Sp}}^{\rm T}(P^\prime)$ and doesn't require a casual prescription. Along the red segment, $P^\prime$ in ${\rm \bold{Sp}}^{\rm T}(P^\prime)$ should have positive imaginary part ${\rm Im} P^\prime\in \Omega_{+}^{p}$, while along the green segment, $P^\prime$ in ${\rm \bold{Sp}}^{\rm S}(-P^\prime)$ should have negative imaginary part ${\rm Im}P^\prime\in\Omega_{-}^{p}$. In principle, every path allowed by analytic continuation should serve the same purpose.

\begin{figure}[ht!]
  \centering
  \includegraphics[width=0.8\linewidth]{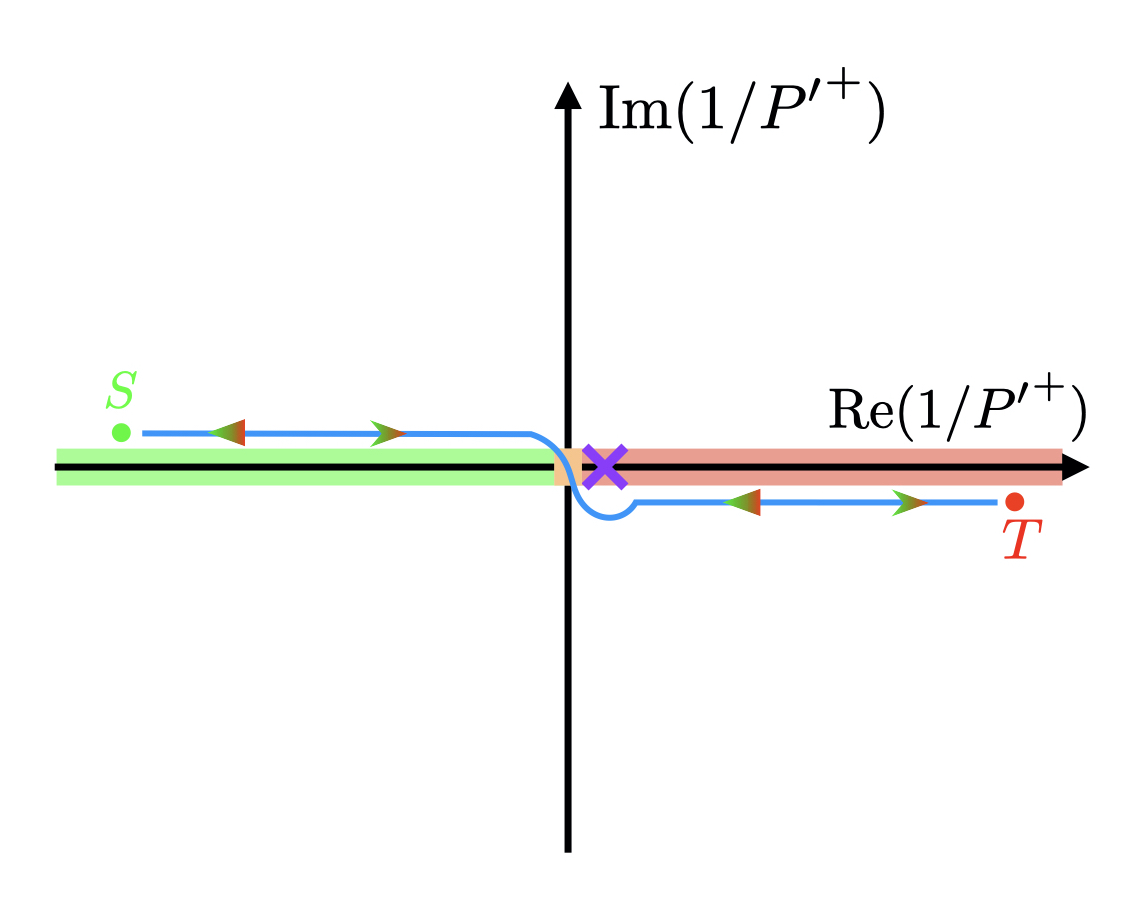}
  \caption{Path of analytic continuation from a time-like point T to a space-like point S or vice versa in the $1/{P^\prime}^{+}$ plane.}
  \label{fig:4}
\end{figure}

The corresponding contour in the complex $1/{P^\prime}^{+}$ plane is depicted schematically in Fig.~(\ref{fig:4}). Note that the orange segment in Fig.~(\ref{fig:3}) can not be simply shown in this plane of single variable, so we abstractly use an orange dot at origin to represent it which allows us to cross the real line analytically. The physical region of time-like process just sits below the positive real line with infinitesimal imaginary part while the physical region of space-like process is just above the negative real line. As illustrated in Fig.~(\ref{fig:4}), the correct path connects $e^{-i0_+}/{P^\prime}^{+}$ and $e^{-i\pi-i0_+}/{{P^\prime}^+}$ for positive ${P^\prime}^+$.

The discussion above determines a unique prescription for the analytic continuation of splitting amplitudes. We define an operator $\underset{\text{T} \to \text{S}}{\cal AC}$ which continue a time-like splitting amplitude from its physical region to a space-like splitting amplitude as
\begin{align}
\underset{\text{T} \to \text{S}}{\cal AC} \circ {\rm \bold{Sp}}_{X_n \bar \imath \leftarrow \bar q}^{\rm T}
(\frac{k_a^+}{P'^+ e^{i0_+}}, \cdots  )
\equiv &\  {\rm \bold{Sp}}_{X_n \bar \imath \leftarrow \bar q}^{\rm T}
(\frac{k_a^+}{P'^+ e^{i (\pi +0_+)}}, \cdots  ) 
\nn
\\
= &\ 
  {\rm \bold{Sp}}_{X_n q \leftarrow i}^{\rm S} (\frac{k_a^+}{P^+ e^{i0_+}}, \cdots) \,.
\end{align}
Similarly for a space-like to time-like continuation, 
\begin{align}
\underset{\text{S} \to \text{T}}
{\cal AC} \circ 
 {\rm \bold{Sp}}_{X_n q \leftarrow i}^{\rm S} (\frac{k_a^+}{P^+ e^{i0_+}}, \cdots) 
\equiv  &\  {\rm \bold{Sp}}_{X_n q \leftarrow i}^{\rm S} (
\frac{k_a^+}{P^+ e^{- i \pi+i0_+}}, \cdots) 
\nn
\\ 
= &\   {\rm \bold{Sp}}_{X_n \bar \imath \leftarrow \bar q}^{\rm T}
(\frac{k_a^+}{P'^+ e^{i0_+}}, \cdots  ) \,.
\end{align}
One can also define analytic continuation operator for complex conjugate amplitudes, 
$\overline{\underset{\text{T} \to \text{S}}{\cal AC}}$, which amounts to perform analytic continuation to amplitudes first, and then take complex conjugate. For a tree-level amplitude, $\underset{\text{T} \to \text{S}}{\cal AC}$ and $\overline{\underset{\text{T} \to \text{S}}{\cal AC}}$ become identical.

\section{Analytic Continuation of TMD Distributions}
\label{sec:ACsplit}

Since TMD distributions are obtained from squared amplitudes, analyticity in external momentum is lost. However, for a subset of contributions to TMD distributions at each perturbative order, it is possible to restore analyticity. We define the holomorphic part of TMD PDFs as~(anti-holomorphic part is simply the conjugate of holomorphic part)
\begin{align}
  {\cal B}_{q/i}^{h} = &\ \int\! d {\rm PS}_{X_n} e^{- i K_\perp \!\cdot b_\perp} 
\delta(K^+ - (1 - x_B) P^+) 
\nn\\
&\ 
e^{- \frac{b_0 \tau}{2} |K^-|} \,
 {\rm \bold{Sp}}_{X_n q^* \leftarrow i}^{\rm S} \left(k_a^+/P^+, \dots \right)
 \frac{\slashed{\bar n}}{2} {\rm \bold{Sp}}_{X_n q^* \leftarrow i}^{\text{S}, (0), *}   \,,
\end{align}
where ${\rm \bold{Sp}}^{\text{S}, (0) ,*}$ is the complex conjugate of tree level splitting amplitude.
We have also inserted a rapidity regulator into the definition of TMD PDFs, which we choose to be exponential regulator $e^{- b_0 \tau |K^-|/2}$~\cite{Li:2016axz,Luo:2019hmp}. The advantage of this regulator is that all the end-point $\delta(1-x)$ terms are absorbed into the soft function~\cite{Li:2016ctv,Billis:2019vxg}, which can be shown to be the same for Drell-Yan, DIS, or $e^+e^-$ processes~\cite{Li:2016ctv,Moult:2018jzp,Li:2020bub}. We emphasize that the results for splitting functions are independent of rapidity regularization. 
In the following, we shall restrict our discussion to $0<x<1$, and show that ${\cal B}_{q/i}^h$ can be analytic continue to ${\cal F}_{\bar \imath/\bar{q}}^h$, and vice versa.

We introduce dimensionless light-cone momentum fraction $y_a = k_a^+/((1-x_B)P^+)$. For $X_n$ consisting of $m$ massless parton, the holomorphic part is
\begin{widetext}
\begin{align}
 {\cal B}_{q/i}^{h,m}(x_B, |P^+|, b_\perp) = 
\int \prod_{a=1}^m  \frac{d^{d-2} \vec{k}_{a,\perp}}{2 (2 \pi)^3} \frac{dy_a}{y_a} 
e^{- i K_\perp \!\cdot b_\perp - \frac{b_0\tau}{2} |K^-|} 
\frac{\delta(\sum_{l=1}^m y_l - 1) }{|1-x_B| |P^+| }
 {\rm \bold{Sp}}_{X_n q^* \leftarrow i}^{\rm S} \left(\frac{y_a(1-x_B)}{e^{i0+}},
 \dots \right)
\frac{\slashed{\bar n}}{2}
{\rm \bold{Sp}}_{X_n q^* \leftarrow i}^{(\text{S}, (0) ,*}   \,,  
\end{align}
where in terms of dimensionless light-cone momentum fraction 
$|K^-| = |\sum_b \vec{k}_{b,\perp}^2 y_b^{-1}|/(|1-x_B| |P^+|)$. The additional $|P^+|$ dependence in the argument of ${\cal B}$ results from rapidity regularization. The analytic continuation reads
\begin{align}
\underset{\text{S} \to \text{T}}{\cal AC} \circ 
 {\cal B}_{q/i}^{h,m} = 
 &\ \int \prod_{a=1}^m  \frac{d^{d-2} \vec{k}_{a,\perp}}{2 (2 \pi)^3} \frac{dy_a}{y_a} 
  e^{- i K_\perp \!\cdot b_\perp - \frac{b_0 \tau}{2} |K^-| } 
\frac{\delta(\sum_{l=1}^m y_l - 1) }{|1-x_B| |P^+| }
 {\rm \bold{Sp}}_{X_n q^* \leftarrow i}^{\rm S} \left(\frac{y_a(1-x_B)}{e^{i(\pi+0_+)}},
 \dots \right) 
\frac{\slashed{\bar n}}{2}
{\rm \bold{Sp}}_{X_n q^* \leftarrow i}^{\text{S}, (0), *} 
\\
= &\ 
\int \prod_{a=1}^m  \frac{d^{d-2} \vec{k}_{a,\perp}}{2 (2 \pi)^3} \frac{dy_a}{y_a} 
  e^{- i K_\perp \!\cdot b_\perp - \frac{b_0 \tau}{2} |K^-| } 
\frac{\delta(\sum_{l=1}^m y_l - 1) }{|1-x_B| |P^+| }
 {\rm \bold{Sp}}_{X_n \bar \imath \leftarrow \bar{q}^*}^{\rm T} \left(\frac{y_a(1-x_B) }{e^{i0_+}},
 \dots \right)
\frac{\slashed{\bar n}}{2}
{\rm \bold{Sp}}_{X_n \bar \imath \leftarrow \bar{q}^*}^{\text{T}, (0), *}  \,.
\label{eq:Bholo}
\end{align}
Note that the analytic continuation operator only acts on the all-order splitting amplitude, as well as the conjugate of tree-level splitting amplitude. We can also write down the holomorphic part of TMD FFs,
\begin{align}
{\cal F}_{\bar \imath/\bar{q}}^{h,m}(x_F, |P'^+|,b_\perp) = &\ x_F^{1-2 \e}   
\int \prod_{a=1}^m  \frac{d^{d-2} \vec{k}_{a,\perp}}{2 (2 \pi)^3} \frac{dy'_a}{y'_a} 
  e^{- i K_\perp \!\cdot b_\perp - \frac{b_0 \tau}{2} |K^-|}
\frac{\delta(\sum_{l=1}^m y'_l - 1) }{|1/x_F - 1| |P'^+| }
 {\rm \bold{Sp}}_{X_n \bar \imath \leftarrow \bar{q}^*}^{\rm T} \left(\frac{y'_a(\frac{1}{x_F}-1)}{e^{i0+}}, \dots \right)
\frac{\slashed{\bar n}}{2} {\rm \bold{Sp}}_{X_n \bar \imath \leftarrow \bar{q}^*}^{\text{T}, (0), *}  \,,
\label{eq:Fholo}
\end{align}
\end{widetext}
where the lightcone momentum fraction is $y'_a = k_a^+/((1/x_F - 1)P'^+)$ and 
$|K^-| = |\sum_b \vec{k}_{b,\perp}^2 y_b'^{-1}|/(|1/x_F-1| |P'^+| )$
we identify the path for analytic continuation,
\begin{align}
(1 - x_B) \to \left( \frac{1}{x_F} - 1\right) e^{i \pi} \,.
\end{align}
The analytic continuation between ${\cal B}^h$ and ${\cal F}^h$ then reads
\begin{gather}
  {\cal F}_{\bar \imath/\bar{q}}^{h,m}(x_F, |P'^+|, b_\perp) = (-1)^{i_F}
 x_F^{1-2 \e}
{\cal B}_{q/i}^{h,m} \left( - \frac{e^{i\pi}}{x_F}, |P'^+|, b_\perp \right) \,,
\label{eq:rulea}
\end{gather}
where $i_F = 1$ if $i$ is a fermion, and $0$ if boson. The minus sign is due to crossing a fermion from initial state to final state.
Similarly for gluon TMD distributions, the analytic continuation reads
\begin{align}
    {\cal F}_{\bar \imath/g}^{h,m}(x_F, |P'^+|, b_\perp) =&\  (-1)^{ 1 + i_F}
 x_F^{1-2 \e}
\nn
\\
&\
\cdot
{\cal B}_{g/i}^{h,m} \left( - \frac{e^{i\pi}}{x_F}, |P'^+|, b_\perp \right) \,,
\label{eq:ruleb}
\end{align}
where the additional minus sign originate from operator definition, and we have suppressed the irrelevant Lorentz indices.
We stress that the analytic continuation is for bare quantities before PDF or FF renormalization.

\begin{figure}[ht!]
  \begin{subfigure}[b]{0.4\linewidth}
    \includegraphics[width=1\textwidth]{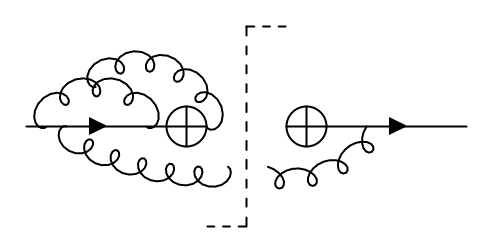}
    \caption{VVR}
\label{fig:5a}
  \end{subfigure}
  \begin{subfigure}[b]{0.4\linewidth}
    \includegraphics[width=1\textwidth]{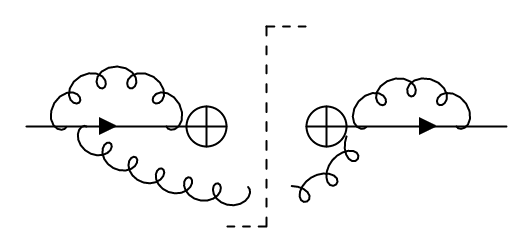}
    \caption{VV$^*$R}
\label{fig:5b}
  \end{subfigure}
\\
  \begin{subfigure}[b]{0.4\linewidth}
    \includegraphics[width=1\textwidth]{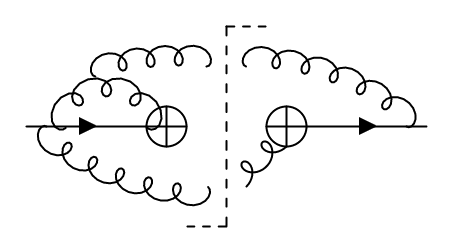}
    \caption{VRR}
\label{fig:5c}
  \end{subfigure}
  \begin{subfigure}[b]{0.4\linewidth}
    \includegraphics[width=1\textwidth]{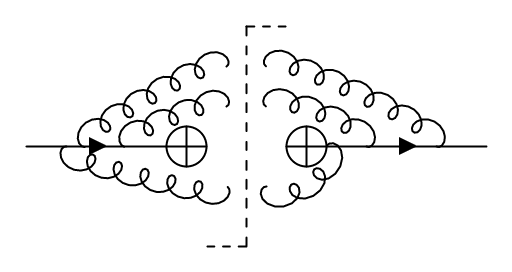}
    \caption{RRR}
\label{fig:5d}
  \end{subfigure}
\caption{Contributions from different partonic channels to TMD PDFs at N$^3$LO.}
\label{fig:5}
\end{figure}

We can now apply the analytic continuation rules in \eqref{eq:rulea} and \eqref{eq:ruleb} to TMD PDFs. At NLO and NNLO, there are only holomorphic and anit- holomorphic contributions. Therefore the analytic continuation rules determine TMD FFs completely. At N$^3$LO, the partonic contributions can be decomposed into triple real~(RRR), double-real virtual~(VRR), double-virtual real~(VVR), and virtual-squared real~(VV$^*$R). The first three contributions are either holomorphic or anti-holomorphic. But the last one, VV$^*$R, mix holomorphic and anti-holomorphic terms, therefore can not be determined from analytic continuation. Since this is a relatively simple contribution, we can calculate it directly using the defining equation in \eqref{eq:amp}. In this way we obtain the bare TMD FFs at N$^3$LO. The results for N$^3$LO TMD FFs will be presented elsewhere. Here we focus on splitting functions. From the single pole terms of bare TMD FFs we extract all the time-like splitting functions through NNLO. Comparing the results with those in the literature, we find full agreement except for the non-diagonal quark-gluon splitting. The discrepancy between our results with those presented in \cite{Almasy:2011eq} can be written as
\begin{gather}
\Delta P_{qg}^{{\rm T},(2)}(x) =  P_{qg}^{{\rm T},(2)}\Big|_{\text{this work}} -  P_{qg}^{\text{T},(2)}\Big|_{\text{\cite{Almasy:2011eq}}}  =
\nn
\\
 \frac{\pi^2}{3} (C_F - C_A) \beta_0 
\left[
-4 + 8 x + x^2 + 6 (1 - 2 x + 2 x^2) \ln x 
 \right] \,,
\end{gather}
where $P_{qg}^{\text{T},(2)}$ is the coefficient of $\alpha_s^3/(4 \pi)^3$ in the off-diagonal singlet splitting matrix, and $\beta_0 = 11 C_A/3 - 2 n_f/3$ is the one-loop  QCD beta function.
In Mellin moment space the discrepancy reads
\begin{gather}
   - \int_0^1 dx\, x^{N-1} \Delta P_{qg}^{{\rm T},(2)} (x)
=  (C_A - C_F) \beta_0 \frac{\pi^2}{3} 
\left( \frac{12}{(N+1)^2}
\right.
\nn
\\
 \left. 
 - \frac{6}{N^2} - \frac{12}{(N+2)^2} - \frac{4}{N} + \frac{8}{N+1} + \frac{1}{N+2}  \right) \,.
\end{gather}
Note that the discrepancy vanishes for $N=2$, as it is completely fixed by the momentum sum rule~\cite{Moch:2007tx}. 
For the convenience of reader we provide the full time-like splitting functions through NNLO as an ancillary file along with the arXiv submission.

\section{Reciprocity Relations in QCD}
\label{sec:recipr-relat-splitt}

With the full space-like and time-like splitting functions, it is interesting to explore yet another relation between them, the so-called reciprocity relation. Reciprocity for tree-level  splitting functions was first proposed by Gribov and Lipatov~\cite{Gribov:1972rt}, which says that $P_{ab}^{\text{S},(0)}(x) = P_{ab}^{\text{T},(0)}(x)$. While the Gribov-Lipatov reciprocity breaks down beyond LO~\cite{Curci:1980uw,Floratos:1981hs}, consideration from small $x$~\cite{Mueller:1982cq,Neill:2020bwv} and large $x$~\cite{Dokshitzer:2005bf,Marchesini:2006ax}, as well as from conformal field theory~\cite{Basso:2006nk,Dokshitzer:2006nm}, suggests a modified form of reciprocity relation exists, at least for the non-singlet. 

Our new results are in the singlet case, which for both space-like and time-like splitting can be written as 
\begin{align}
\widehat{P}(x,\alpha_s) = 
  \begin{pmatrix}
    \widetilde{P}_{qq} & 2 n_f P_{qg}
\\
    P_{gq} & P_{gg} 
  \end{pmatrix} \,,
\end{align}
where
\begin{equation}
 \widetilde{P}_{qq} = P_{qq} + P_{\bar{q} q} + (n_f-1)(P_{q'q}+P_{\bar{q}'q})  \,.
\end{equation}
For time-like splitting, the $P_{ij}^{\text{T}}$ can be found in the ancillary file through NNLO. 
It is also convenient to introduce the Mellin moment of singlet splitting,
\begin{align}
\widehat{\gamma}(N,\alpha_s) = - \int_0^1 dx \, x^{N-1} \widehat{P}(x,\alpha_s)  \,,
\end{align}
and the associate eigenvalues, 
\begin{equation}
\gamma_\pm = 
\frac{1}{2}( \pm \sqrt{(\text{tr}\widehat{\gamma})^2 - 4 \text{det} \widehat{\gamma} }  + \text{tr}\widehat{\gamma} ) \,.
\end{equation}
An important motivation for reciprocity relation in singlet comes from the evolution equation for jet functions in energy correlators~\cite{Dixon:2019uzg,Chen:2020vvp},
\begin{align}
\frac{d \vec{J} (\ln \frac{x_L Q^2}{\mu^2})}{d \ln\mu^2}
= \int_0^1 \! dy\, y^N   
\vec{J} (\ln \frac{x_L y^2 Q^2}{\mu^2}) \cdot \widehat{P}^{\rm T} (y,\alpha_s) \,,
\label{eq:jetevo}
\end{align}
where $x_L$ measures the size of $N$ tagged particles in a jet. Note that this is an non-local evolution equation. For fixed coupling, one can write down for \eqref{eq:jetevo} a completely local, power-law solution for $\vec{J}$, with the power-law exponent given by $\gamma_{\pm}^{\text{T}}$ evaluated at a shift $N$. Based on this consideration, we propose the following
reciprocity relations for the singlet splitting with running coupling, 
\begin{align}
 2 \gamma_\pm^{\rm S}(N, \alpha_s) = &\ 2 \gamma_\pm^{\rm T} (N + 2 \gamma_\pm^{\rm S} (N, \alpha_s), \alpha_s) \,,
\label{eq:reciprocity1}
\\
 2 \gamma_\pm^{\rm T}(N, \alpha_s) = &\ 2 \gamma_\pm^{\rm S} (N - 2 \gamma_\pm^{\rm T} (N, \alpha_s), \alpha_s) \,.
\label{eq:reciprocity2}
\end{align}
The two relations \eqref{eq:reciprocity1} and \eqref{eq:reciprocity2} are not independent. We have verified \eqref{eq:reciprocity1} and \eqref{eq:reciprocity2} through NNLO~($\alpha_s^3$) using the newly determined time-like singlet splitting functions. On the other hand, this relation is violated should we use the $P_{qg}^{\text{T},(2)}$ from \cite{Almasy:2011eq}.  We stress that the reciprocity relation is for arbitrary $N$, and therefore hints at hidden relation between space-like and time-like process beyond small $x$ and large $x$.

\section{Conclusion}
\label{sec:conclusion}

We have provided a clean theoretical understanding of analytic continuation for TMD distributions and splitting functions using SCET. Employing the analytic continuation rules for holomorphic and anti-holomorphic contributions to TMD distributions, we determined the time-like splitting functions in QCD through NNLO. For the eigenvalues of the singlet splitting matrix, we propose an all-order reciprocity relation, valid for arbitrary $N$. We verified this relation through NNLO using the newly determined time-like singlet splitting functions. We leave a deeper understanding of the reciprocity relation to future work.

\begin{acknowledgments}
We thank Lance Dixon, Yi-Bei Li, Ming-xing Luo, Ian Moult, and Hua-Sheng Shao for helpful discussion.
This work was supported in part by National Natural Science Foundation of China under contract No.~11975200, and the Zhejiang University Fundamental Research Funds for the Central Universities (2019QNA3005).
\end{acknowledgments}

\bibliography{crossing-reciprocity.bib}{}
\bibliographystyle{apsrev4-1}


\end{document}